\documentclass[12pt]{article}

\begin{document}

\input amssym.tex

\title{The free Dirac spinors of the spin basis on the de Sitter expanding universe}

\author{Ion I. Cot\u aescu \thanks{E-mail:~~cota@physics.uvt.ro}\\
{\small \it West University of Timi\c soara,}\\
{\small \it V. Parvan Ave. 4 RO-300223 Timi\c soara,  Romania}}

\maketitle

\begin{abstract}
It is shown that on the de Sitter space-time the global behavior of
the free Dirac spinors in momentum representation is determined by
several phases factors which are functions of momentum with special
properties. Such suitable phase functions can be chosen for writing
down the free Dirac quantum modes of the spin basis that are
well-defined even for the particles at rest in the moving local
charts where the modes of the helicity basis remain undefined. Under
quantization these modes lead to a basis in which the one-particle
operators keep their usual forms apart from the energy operator
which lays out a specific term which depend on the concrete phase
function one uses.

Pacs: 04.62.+v
\end{abstract}


\newpage

The quantum fields with spin half on curved manifolds are less
studied because of the asperities of the gauge-covariant theory in
non-holonomic frames where these fields can be defined \cite{SW,BD}.
For this reason we are left with some delicate questions to address
even in the case of the de Sitter space-time where the free field
equations can be analytically solved. One of them is related to the
definition of the quantum modes in momentum representation of the
particles which stay at rest in a given local chart. We discuss here
this problem focusing on the Dirac field minimally coupled to
gravity in (co)moving charts of the de Sitter expanding universe
\cite{BD}.

The first solutions of the free Dirac equation in a moving charts
with proper time and spherical coordinates were derived by Shishkin
\cite{SHI} and normalized in Ref. \cite{C4}. We derived other
solutions of this equation but in moving charts with Cartesian
coordinates where we considered the helicity basis in momentum
representation \cite{CD1}. These solutions are well-normalized and
satisfy the usual completeness relations. Notice that there are
other attempts to write down solutions in momentum representation
but these are not correctly normalized \cite{BF,GF}.

A specific problem arising in the case of the helicity bases is that
the solutions in the rest frames remain undefined since the helicity
does not make sense for vanishing momentum. For this reason it is
worth analyzing other spinor bases in which the polarization can be
defined even for the particles at rest, as in the case of the spin
basis \cite{BDR} considered already in \cite{BF,GF}. The technical
problem here is that the limits of the Dirac spinors for vanishing
momentum are not trivial because of some undefined phase factors of
the functions giving the time modulation of these spinors. On the
other hand, the Dirac equation in the rest frames has well-defined
solutions which can be interpreted as rest spinors. Under such
circumstances, we must look for suitable momentum-dependent phase
factors of the Dirac spinors so that their limits for vanishing
momentum should be just the mentioned rest spinors. In what follows
we present this procedure.

Let us consider $(M,g)$ be the de Sitter expanding universe of
radius $\frac{1}{\omega}$ where the notation $\omega$ stands for its
Hubble constant. We choose the moving chart $\{t,\vec{x}\}$ of the
{\em conformal} time, $t\in (-\infty,0]$, Cartesian coordinates and
the line element
\begin{equation}
ds^{2} =\frac{1}{(\omega t)^2}\,\left(dt^{2}- d\vec{x}\cdot
d\vec{x}\right)\,,
\end{equation}
which covers the expanding part of the de Sitter manifold. In
addition, we use the non-holonomic frames defined by the tetrad
fields which have only diagonal components,
\begin{equation}\label{tt}
e^{0}_{0}=-\omega t\,, \quad e^{i}_{j}=-\delta^{i}_{j}\,\omega t \,,\quad \hat
e^{0}_{0}=-\frac{1}{\omega t}\,, \quad \hat e^{i}_{j}=-\delta^{i}_{j}\,
\frac{1}{\omega t}\,.
\end{equation}
In this tetrad-gauge, the free Dirac equation, $E_D \psi =m\psi$,
which is governed by the Dirac operator
\begin{equation}\label{ED}
E_D=-i\omega t\left(\gamma^0\partial_{t}+\gamma^i\partial_i\right)
+\frac{3i\omega}{2}\gamma^{0}\,,
\end{equation}
can be analytically solved obtaining the momentum and energy bases
with correct normalization factors \cite{CD1,CD2}.

The plane wave solutions of the momentum basis and arbitrary
polarization $\sigma$ can be derived as in Ref. \cite{CD1} starting
with the mode expansion
\begin{equation}\label{psiab}
\psi(t,\vec{x})=\int d^3 p \sum_{\sigma}\left[U_{\vec{p},\sigma}(x)
a(\vec{p},\sigma)+V_{\vec{p}, \sigma}(x){a^c}^{\dagger}(\vec{p},
\sigma) \right]\,,
\end{equation}
and solving then the Dirac equation in the standard representation
of the Dirac matrices (with diagonal $\gamma^0$). Thus we obtain the
particle and antiparticle fundamental solutions,
\begin{eqnarray}
U_{\vec{p},\sigma}(t,\vec{x}\,)&=& i N (\omega t)^2\left(
\begin{array}{c}
\,e^{\pi \mu/2}H^{(1)}_{\nu_{-}}(-p t) \,
\xi_{\sigma}\\
 e^{-\pi \mu/2}H^{(1)}_{\nu_{+}}(-p t) \,
 \frac{\vec{p}\cdot\vec{\sigma}}{p}\,\xi_{\sigma}
\end{array}\right)
e^{i\vec{p}\cdot\vec{x}}\label{Ups}\\
V_{\vec{p},\sigma}(t,\vec{x}\,)&=&-i N (\omega t)^2 \left(
\begin{array}{c}
e^{-\pi \mu/2}H^{(2)}_{\nu_{-}}(-p t)\,
\frac{\vec{p}\cdot\vec{\sigma}}{p}\,\eta_{\sigma}\\
e^{\pi \mu/2}H^{(2)}_{\nu_{+}}(-p t) \,\eta_{\sigma}
\end{array}\right)
e^{-i\vec{p}\cdot\vec{x}}\,,\label{Vps}
\end{eqnarray}
where $p=|\vec{p}|$, $H_{\nu_{\pm}}^{(1,2)}$ are the Hankel
functions of indices $\nu_{\pm}=\frac{1}{2}\pm i\mu$, with
$\mu=\frac{m}{\omega}$, while
\begin{equation}
N=\frac{1}{2(2\pi)^{3/2}}\sqrt{\frac{\pi p}{\omega}}\,,
\end{equation}
is the normalization constant which assures the good
orthonormalization and completeness properties \cite{CD1}. These
properties do not depend on the concrete choice of the Pauli spinors
$\xi_{\sigma}$ and $\eta_{\sigma}= i\sigma_2 (\xi_{\sigma})^{*}$ if
these are correctly normalized as
$\xi^+_{\sigma}\xi_{\sigma'}=\eta^+_{\sigma}\eta_{\sigma'}=\delta_{\sigma\sigma'}$.
In Ref. \cite{CD1} we used the Pauli spinors of the {\em helicity}
basis in which the direction of the spin projection is just that of
the momentum $\vec{p}$. However, we can project the spin on an
arbitrary direction, independent on $\vec{p}$, as in the case of the
{\em spin} basis \cite{BDR} where $\xi_{\frac{1}{2}}=(1,0)^T$ and
$\xi_{-\frac{1}{2}}=(0,1)^T$ for particles and
$\eta_{\frac{1}{2}}=(0,-1)^T$ and $\eta_{-\frac{1}{2}}=(1,0)^T$ for
antiparticles. We consider here that the Pauli spinors of the
solutions (\ref{Ups}) and (\ref{Vps}) are those of the spin basis
since these do make sense even in the {\em natural rest frame} where
$\vec{p}=0$.

Unfortunately, these solutions remain undefined for $\vec{p}= 0$
since the phases of the Hankel functions are undefined at this point
(behaving as $0^{\pm i}$). On the other hand, in the rest frame, the
Dirac equation has well-defined solutions of the spin basis  that
read
\begin{eqnarray}
\tilde U_{0,\sigma}(t)&=&  N' (-\omega t)^{\frac{i}{\omega} E_0^+
}\left(
\begin{array}{c}
\xi_{\sigma}\\
 0
\end{array}\right)\,,\label{cucu}\\
\tilde V_{0,\sigma}(t)&=& N'' (-\omega t)^{\frac{i}{\omega} E_0^-}
\left(
\begin{array}{c}
0\\
\eta_{\sigma}
\end{array}\right)\label{mucu}\,.
\end{eqnarray}
We note that in this frame the polarizations $\sigma=\pm
\frac{1}{2}$ represent the spin projections on the third axis of the
non-holonomic frame which in our gauge (\ref{tt}) is parallel to
that of the natural rest frame. The quantities
$E_0^{\pm}=\pm\,m-\frac{3i\omega}{2}$ are the particle/antiparticle
rest energies whose last term is due to the decay produced by the
expansion of the de Sitter expanding universe \cite{CCC}. The
normalization constants $N'$ and $N''$ are not yet specified.

Hereby a delicate problem is arising, namely:\\

\noindent{\em How the phases of the spinors (\ref{Ups}) and
(\ref{Vps}) must be changed in order to obtain fundamental solutions
continuous on the whole space ${\Bbb R}_p^3$, including the point $\vec{p}=0$}\\

\noindent Obviously, we must solve this problem if we want to
understand what happens with the particles at rest in the chart
$\{t,\vec{x}\}$.

The solution is not trivial even though the spinors (\ref{Ups}) and
(\ref{Vps}) are defined up to arbitrary phase factors which do not
affect the relativistic scalar product. This is because in the de
Sitter case there are conserved quantities whose forms are
determined by the choice of some phase factors depending on
$\vec{p}$. Let us explain this mechanism concentrating on the
conserved one-particle operators associated to the de Sitter
symmetries \cite{CD1}.

Under canonical quantization, the particle $(a,a^{\dagger})$ and
antiparticle $(a^c,{a^c}^{\dagger})$ field operators satisfy the
non-vanishing anti-commutators \cite{CD1}
\begin{equation}\label{acom} \{a(\vec{p},\sigma),
a^{\dagger}({\vec{p}}^{\,\prime},\sigma^{\prime})\}= \{{a^c}(\vec{p},\sigma),
{a^c}^{\dagger}({\vec{p}}^{\,\prime},\sigma^{\prime})\}=
\delta_{\sigma\sigma^{\prime}}\delta^3 (\vec{p}-{\vec{p}}^{\,\prime})\,.
\end{equation}
The principal one-particle operators  are the (electric) charge
operator
\begin{equation}\label{QQ}
{\cal Q}= \int d^3 p  \sum_\sigma
\left[a^{\dagger}(\vec{p},\sigma)a(\vec{p},\sigma)
-{a^c}^{\dagger}(\vec{p},\sigma){a^c}(\vec{p},\sigma)\right]
\end{equation}
and the components of the momentum operator,
\begin{equation}\label{PP}
{\cal P}^i= \int d^3 p\, p^i\sum_{\sigma}
\left[a^{\dagger}(\vec{p},\sigma)a(\vec{p},\sigma)
+{a^c}^{\dagger}(\vec{p},\sigma){a^c}(\vec{p},\sigma)\right]\,,
\end{equation}
which are diagonal in the momentum basis. The polarization operator
is also diagonal but its form depends on the direction along which
one measures the spin projection. The energy operator, ${\cal H}$,
is conserved but is not diagonal in the momentum basis since it does
not commute with ${\cal P}^i$. Nevertheless, this may be written in
momentum representation as \cite{CD1},
\begin{equation}\label{energy}
{\cal H}={\cal H}[a,a^c]=\frac{i\omega}{2}\int d^3
p\,p^i\sum_{\sigma}\left[
a^{\dagger}(\vec{p},\sigma)\stackrel{\leftrightarrow}{\partial}_{p^{i}}
a(\vec{p},\sigma) +
{a^c}^{\dagger}(\vec{p},\sigma)\stackrel{\leftrightarrow}{\partial}_{p^{i}}
{a^c}(\vec{p},\sigma) \right]
\end{equation}
where we use the notation
$f\stackrel{\leftrightarrow}{\partial}h=f\partial h -(\partial f)
h$.

We arrive now at the tool able to solve our problem. We have shown
that there exists momentum-dependent $U(1)$ phase transformations
which play a central role in interpreting the energy operator
\cite{CD1}. These transform simultaneously the spinors,
\begin{eqnarray}
U_{\vec{p},\sigma}(t,\vec{x}\,)&\to&\tilde
U_{\vec{p},\sigma}(t,\vec{x}\,)=
e^{-i\chi(\vec{p})}U_{\vec{p},\sigma}(t,\vec{x}\,)\,,\\
V_{\vec{p},\sigma}(t,\vec{x}\,)&\to&\tilde
V_{\vec{p},\sigma}(t,\vec{x}\,)=
e^{i\chi(\vec{p})}V_{\vec{p},\sigma}(t,\vec{x}\,)\,,
\end{eqnarray}
and the field operators,
\begin{eqnarray}\label{gaugeab}
a(\vec{p},\sigma)&\to&\tilde a(\vec{p},\sigma)= e^{i\chi(\vec{p})}a(\vec{p},\sigma)\,,\\
a^c(\vec{p},\sigma)&\to&\tilde a^c(\vec{p},\sigma)=
e^{i\chi(\vec{p})}a^c(\vec{p},\sigma)\,,
\end{eqnarray}
with phase factors depending on {\em real} functions
$\chi(\vec{p})$. Any such transformation preserves the form of the
field (\ref{psiab}) and the operators ${\cal Q}$ and ${\cal P}_i$
but changes the form of the energy operator as
\begin{equation}
{\cal H}={\cal H}[\tilde a,\tilde a^c]+{\cal H}_{\chi}[\tilde
a,\tilde a^c]\,,
\end{equation}
where $H[\tilde a,\tilde a^c]$ has the same form as in equation
(\ref{energy}) and
\begin{equation}\label{Hchi}
{\cal H}_{\chi}[\tilde a,\tilde a^c]=  \omega\int d^3 p\, [p^i
\partial_{p^i}\chi(\vec{p})] \sum_\sigma
\left[\tilde a^{\dagger}(\vec{p},\sigma)\tilde a(\vec{p},\sigma)
+{\tilde a}^{c\,\dagger}(\vec{p},\sigma){\tilde
a^c}(\vec{p},\sigma)\right]\,.
\end{equation}
This phenomenon is new since in the flat case there are no similar
transformations able to change the expressions of the one-particle
operators.

With these preparations we can solve our problem assuming that the
phase function $\chi(\vec{p})$ of the transformed spinors must be
fixed so that the limits of these spinors at $\vec{p}= 0$  do make
sense and coincide to the rest solutions (\ref{cucu}) and
(\ref{mucu}). Taking into account that $N$ includes the factor
$\left(\frac{p}{\omega}\right)^{\frac{1}{2}}$ and using the limits
of the Hankel functions,
\begin{equation}
\lim_{{x}\to 0}x^{\nu} H^{(1)}_{\nu}(\alpha x)=-\lim_{{x}\to
0}x^{\nu} H^{(2)}_{\nu}( \alpha
x)=\frac{1}{i\pi}\left(\frac{2}{\alpha}\right)^{\nu}\Gamma(\nu)\,,
\end{equation}
that hold for $\Re \nu >0$, we find that:\\

\noindent {\em The transformed spinors $\tilde
U_{\vec{p},\sigma}(t,\vec{x}\,)$ and $\tilde
V_{\vec{p},\sigma}(t,\vec{x}\,)$ have well-defined limits at
$\vec{p}= 0$  if their phase function satisfies}
\begin{equation}\label{chichi}
\lim_{\vec{p}\to
0}\left[\chi(\vec{p})-\mu\ln\left(\frac{p}{\omega}\right)\right]=0\,.
\end{equation}

\noindent With such phases,  the transformed spinors  become
continuous on the whole momentum space since the indetermination of
$\frac{\vec{p}\cdot\vec{\sigma}}{p}$ at $\vec{p}=0$ is rather
apparently as long as such terms do not appear in the Dirac equation
written for $\vec{p}=0$. Moreover, we can determine the definitive
form of the rest spinors (\ref{cucu}) and (\ref{mucu}) assuming
that,
\begin{equation}
\lim_{\vec{p}\to 0}\tilde U_{\vec{p},\sigma}= \tilde U_{0,\sigma}\,,
\quad \lim_{\vec{p}\to 0}\tilde V_{\vec{p},\sigma}= \tilde
V_{0,\sigma}\,.
\end{equation}
The resulting normalization constants,
\begin{equation}
N'= (N'')^*=\frac{e^{\frac{\pi\mu}{2}-i\mu\ln
2}}{(2\pi)^2}\,\Gamma\left(\textstyle{\frac{1}{2}}-i\mu\right)\,,
\end{equation}
satisfy $|N'|=|N''|=
(2\pi)^{-\frac{3}{2}}\left(1+e^{-2\pi\mu}\right)^{-\frac{1}{2}}$.

Hence our problem is completely solved. It is remarkable that there
are many phase functions obeying the condition (\ref{chichi}). The
simplest particular case is of the choice
\begin{equation}
\chi(\vec{p})=\mu\ln\left(\frac{p}{\omega}\right)
\end{equation}
for which the term (\ref{Hchi}) takes the form ${\cal
H}_{\chi}[\tilde a,\tilde a^c]=m {\cal N}$ where
\begin{equation}
{\cal N}=\int d^3 p\,\sum_{\sigma}\left[ \tilde
a^{\dagger}(\vec{p},\sigma) \tilde a(\vec{p},\sigma) + {\tilde
a}^{c\,\dagger}(\vec{p},\sigma) {\tilde a}^c(\vec{p},\sigma)
\right]\,,
\end{equation}
is the operator of the number of particles. In other words,  this
phase fixing separates just the real part of the {rest energy} which
is the same as in special relativity. A more interesting separation
can be done by choosing the phase function
\begin{equation}
\chi(\vec{p})=\left(\mu^2+\frac{p^2}{\omega^2}\right)^{\frac{1}{2}}-\mu\,{\rm
arctanh}\,\mu
\left(\mu^2+\frac{p^2}{\omega^2}\right)^{-\frac{1}{2}}+\mu\left( \ln
2\mu-1\right)
\end{equation}
which satisfies the condition (\ref{chichi}) giving rise to the term
\begin{equation}\label{Hchi1}
{\cal H}_{\chi}[\tilde a,\tilde a^c]=  \int d^3
p\,\sqrt{m^2+{\vec{p}}^{\,2}} \sum_\sigma \left[\tilde
a^{\dagger}(\vec{p},\sigma)\tilde a(\vec{p},\sigma) +{\tilde
a}^{c\,\dagger}(\vec{p},\sigma){\tilde
a^c}(\vec{p},\sigma)\right]\,,
\end{equation}
representing  the energy operator of special relativity (when
$\omega\to 0$). This result could be useful for studying the flat
limit of our theory which is faced with serious mathematical
difficulties arising from the fact that one can not evaluate the
limits for $\omega\to 0$ of the Hankel functions of the spinors
(\ref{Ups}) and (\ref{Vps}) by using analytical methods \cite{CD1}.

The conclusion is that the phase functions can be chosen in order to
write down the continuous Dirac quantum modes of the spin basis
which are well-defined on the whole momentum space. In this basis
and the gauge (\ref{tt}) the polarization is measured with respect
to the third axis of the natural rest frame. Obviously, this
direction can be changed at any time by changing the gauge
(\ref{tt}) rotating the axes of the local frames. These results will
allow one to calculate the amplitudes of the quantum transitions
involving fermions in gravitational fields whose polarizations are
measured with respect to the fixed directions of the experimental
devices. Of a special interest could be the study of the particle
creation on the de Sitter expanding universe using perturbations in
a further quantum field theory which may complete the results
obtained so far using the WKB method \cite{WKB,WKB1,WKB2}.

Finally we observe that  the method presented here works only for
$\Re \nu \not =0$. This is useless for the scalar \cite{Cs} or
massive vector \cite{Cv} fields minimally coupled to the de Sitter
gravity since in both these cases $\Re \nu=0$. This means that the
limits at $\vec{p}= 0$ of the mode functions of these fields are
undefined as long as these remain minimally coupled to the de Sitter
gravity. This could be an argument for considering other types of
couplings of the scalar and massive vector fields.

\subsection*{Acknowledgements}

This work is partially supported by the ICTP-SEENET-MTP grant PRJ-09
in frame of the SEENET-MTP Network G. S. Dj..

\end{document}